# Computer Modelling of 3D Geological Surface

Kodge B. G.
Department of Computer Science,
S. V. College, Udgir, District Latur,
Maharashtra state, India
kodgebg@hotmail.com

Hiremath P. S.
Department of Computer Science,
Gulbarga University, Gulbarga
Karnataka state, India
Hiremathps53@yahoo.com

*Abstract*— **The geological surveying presently uses methods and tools for the computer modeling of 3D-structures of the geographical subsurface and geotechnical characterization as well as the application of geoinformation systems for management and analysis of spatial data, and their cartographic presentation. The objectives of this paper are to present a 3D geological surface model of Latur district in Maharashtra state of India. This study is undertaken through the several processes which are discussed in this paper to generate and visualize the automated 3D geological surface model of a projected area.**

## I. INTRODUCTION

Traditional geological maps which illustrate the distribution and orientation of geological structures and materials on a two-dimensional (2D) ground surface are no longer sufficient for the storing, displaying, and analysing of geological information. It is also difficult and expensive to update traditional maps that cover large areas. Many kinds of raster and vector based models for describing, modelling, and visualizing 3D spatial data have been developed. At the mean time, with the fast development of sensor techniques and computer methods, several types of airborne or close range laser scanners are available for acquisition of 3D surface data in real or very fast time. A few more type of digital photogrammetry workstations are also available for semi-automatic interpretation of the complicated man made 3D surfaces. However due to image noises and limited resolution of current laser range data, so many existing techniques still need to be extended to fit real application.

This paper presents a fast and efficient method to automate the generation of 3D geological surfaces from 2D geological maps. The method was designed to meet the requirement in creating a three-dimensional (3D) geologic map model of Latur district in Maharashtra state of India. The LULC (Land Use and Land Cover) database [11] of National Remote Sensing Centre, ISRO, India, for Latur district has been used for visualization experiments. The elevation data pertaining to Latur district is obtained from USGS (United State Geological Survey) Seamless server database [10] of United States and is used for digital elevation modelling (DEM) experiments.

## II. STUDY AREA

Latur District is in the south-eastern part of the Maharashtra state in India. It is well known for its Quality of Education, Administration, food grain trade and oil mills. Latur district has an ancient historical background. The King 'Amoghvarsha' of Rashtrakutas developed the Latur city, originally the native place of the Rashtrakutas. The Rashtrakutas who succeeded the Chalukyas of Badami in 753 A.D called themselves the residents of Lattalut. Latur is a major city and district in Maharashtra state of India. It is well known for its quality of education, administration, food grain trade and oil mills. The district is divided into three sub-divisions and 10 talukas (sub-districts) [1].

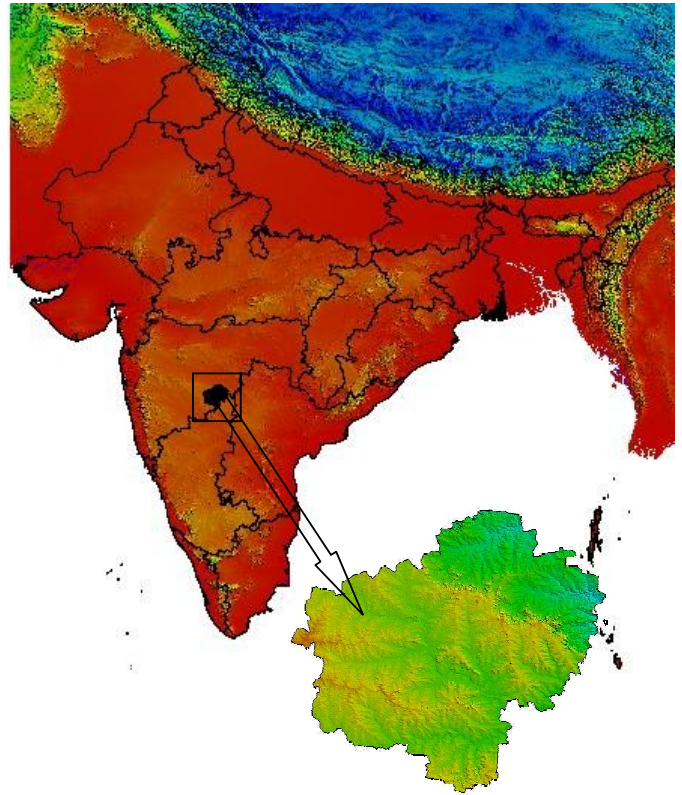

Figure 1. A false color composite imagery of India acquired by SPOT & IKONOS, the location of Latur district (Courtesy NRSA Hyd.).

Latur is located at 18°24′N 76°35′E / 18.4°N 76.58°E / 18.4; 76.58 as shown in Fig.1. It has an average elevation of 631 meters (2070 feet). It is situated 636 meter above mean sea level. The district is situated on Maharashtra-Karnataka boundary. On the eastern side of the Latur is Bidar district of Karnataka, whereas Nanded is on the Northeast, Parbhani district on the northern side, Beed on the Northwest and Osmanabad on the western and southern side. The entire district of Latur is situated on the Balaghat plateau, 540 to 638 meters from the mean sea level.

### III. AUTOMATED 3D SURFACE MODEL

3D geological information systems provide a means to capture, model, manipulate, retrieve, analyse, and present geological situations. Traditional geological maps which illustrate the distribution and orientation of geological materials and structures on a 2D ground surfaces provide vast amounts of raw data. It is thus vital to develop a set of intelligent maps that shows features of geological formations and their relationships[2].

#### A. Digital Elevation Model of Latur district

DEM is a representation of the terrain surface by coordinates and numerical descriptions of altitude. DEM is easy to store and manipulate, and it gives a smoother, more natural appearance of derived terrain features. Therefore, the created DEM is the foundation of 3D geological maps when the z-coordinates of the vertices of geological formations can be interpolated. The data consists of 4 topographical map sheets, with 3D coordinates of terrain, contour lines, and other information. The maps are in GEOTIFF format at a scale of 1:150000 (Fig.2). These DEMs were then integrated into a whole DEM of Latur using a DEM Global Mapper. The final gridded DEM data with 5-metre intervals for Latur district was obtained (Fig.2). The file size is about 4.83MB.

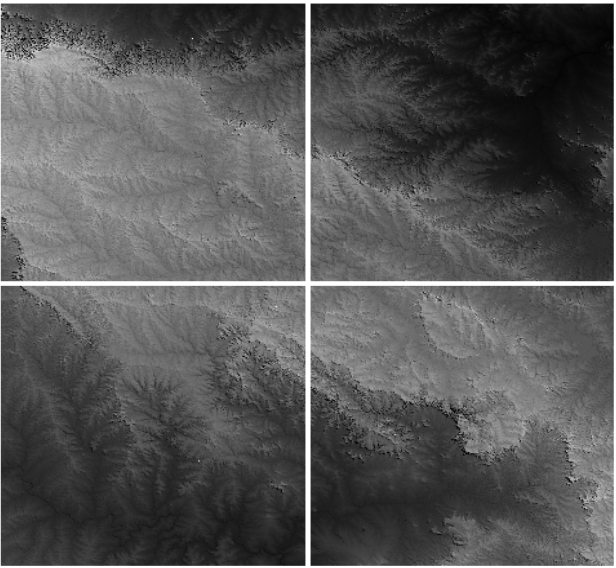

Figure 2. Tiled DEM of Latur District (courtesy USGS).

#### B. Cropping DEMs using Latur district base map shape file.

After integrating DEMs tiles, the next process is to extract (crop) the required region of Latur district from integrated DEMs using the latur district base map shape file. For this process, we use the software GLOBAL MAPPER 11v to crop the DEMs with only required region's terrain data. The remaining area is considered as null data as shown in Fig.3.

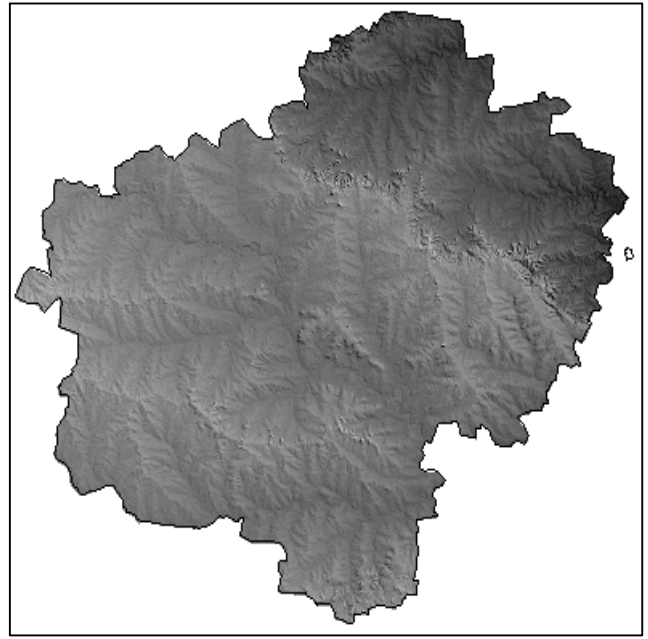

Figure 3. Cropped DEM using Latur district base map.

#### C. Accessing and concatenating DEMs in MATLAB

After the successful cropping of all the DEM data sheets (tiles), we import them in MATLAB for further processes. The DEMs can be converted in to DTED (Digital Terrain Elevation Data) version 0,1,2.. any format, and import them in MATLAB. The DTED0 files have 120-by-120 points. DTED1 files have 1201-by-1201. The edges of adjacent tiles have redundant records.

Acquiring all the data sheets with their specified location (projection) and sequence of data sheets are very important here.

Concatenation of the DEM tiles with respect to their locations needs horizontal and vertical concatenation.

*1) Horizontal Concatenation*

First, we concatenate the matrices of top-left and top-right tiles (Fig.2), i.e. Horizontal concatenation.

$$H1 = TL \text{ (horzcat) } TR . \qquad (1)$$

where H1 is a concatenated matrix of top-left (TL) and top-right (TR) matrices.

Next, we concatenate the matrices of Bottom-Left and Bottom-Right tiles, i.e. again Horizontal concatenation.

$$H2 = BL \text{ (horzcat) } BR \quad (2)$$

where H2 is a concatenated matrix of Bottom-left (BL) and Bottom-right (BR) matrices.

*2) Vertical Concatenation*

Next, we need to concatenate H1 and H2 matrices vertically, i.e.

$$H = H1 \text{ (vertcat) } H2 \quad (3)$$

where H is a complete concatenated matrix of H1 and H2.

*D. Visualizing 3D geographical surface model*

A workflow was chosen, on the one hand, by applying GIS methods using ESRI shape files and global mapper software for data acquisition, maintenance, and presentation and on the other hand, by applying three-dimensional spatial modelling with a interactive 3D modelling in MATLAB. Based on Non-Uniform Rational data, any geometric shape can be modelled. Besides surfaces of the different engineering geological units, solids using boundary representation techniques were modelled [3]. In MATLAB it is one of the easiest way to visualize the well defined projected data sets in 3D view using mathematical functions surf() and mesh(). To visualize the acquired projected data set over a rectangular region, we need to create colored parametric surfaces specified by X, Y, and Z, with color specified by Z.

A parametric surface is parameterized by two independent variables, i and j, which vary continuously over a rectangle; for example, $1<=i<=m$ and $1<=j<=n$. The three functions $x(i,j)$, $y(i,j)$, and $z(i,j)$ specify the surface. When i and j are integer values, they define a rectangular grid with integer grid points. The functions $x(i,j)$, $y(i,j)$, and $z(i,j)$ become three m-by-n matrices, X, Y, and Z. Surface color is a fourth function, $c(i,j)$, denoted by matrix C. Each point in the rectangular grid can be thought of as connected to its four nearest neighbours [6].

$$\begin{array}{c} i\text{-}1,j \\ | \\ i,j\text{-}1 - i,j - i,j\text{+}1 \\ | \\ i\text{+}1,j \end{array} \quad (4)$$

Surface color can be specified in two different ways: at the vertices or at the centers of each patch. In this general setting, the surface need not be a single-valued function of x and y. Moreover, the four-sided surface patches need not be planar. For example, one can have surfaces defined in polar, cylindrical, and spherical coordinate systems [8].

The shading function sets the shading. If the shading is interpolates, C must be of the same size as X, Y, and Z; it specifies the colors at the vertices. The color within a surface patch is a bilinear function of the local coordinates. If the shading is faceted (the default) or flat, $C(i,j)$ specifies the constant color in the surface patch:

$$\begin{array}{ccc} (i,j) & - & (i,j+1) \\ | & C(i,j) & | \\ (i+1,j) & - & (i+1,j+1) \end{array} \quad (5)$$

In this case, C can be the same size as X, Y, and Z and its last row and column are ignored. Alternatively, its row and column dimensions can be one less than those of X, Y, and Z.

*E. Assigning axes to 3D model*

MATLAB automatically creates an axes, if one does not already exist, when you issue a command that creates a graph, but the default axes assigned by MATLAB doesn't match with real coordinate systems of this projected area.

This existing model is built with 3 axes data x, y and z respectively. The X and Y axis represents the latitude and longitude values for this model i.e.

UPPER LEFT X=76.2076079218
UPPER LEFT Y=18.8385493143
LOWER RIGHT X=77.2934412815
LOWER RIGHT Y=17.8677159574

WEST LONGITUDE=76° 12' 27.3885" E
NORTH LATITUDE=18° 50' 18.7775" N
EAST LONGITUDE=77° 17' 36.3886" E
SOUTH LATITUDE=17° 52' 3.7774" N

The above shown values are associated with all four tiles of DTED files. The Z axis itself represents the terrain (height) values of ground surface objects. Here in this model the elevation data is assigned in feet scale format i.e. 0 to 3000 feets.

IV. RESULTS

With reference to the processes discussed above, the 3D visualization experimental results are shown in the Figs. 4, 5 and 6 for 3D model of Latur district geological surface.

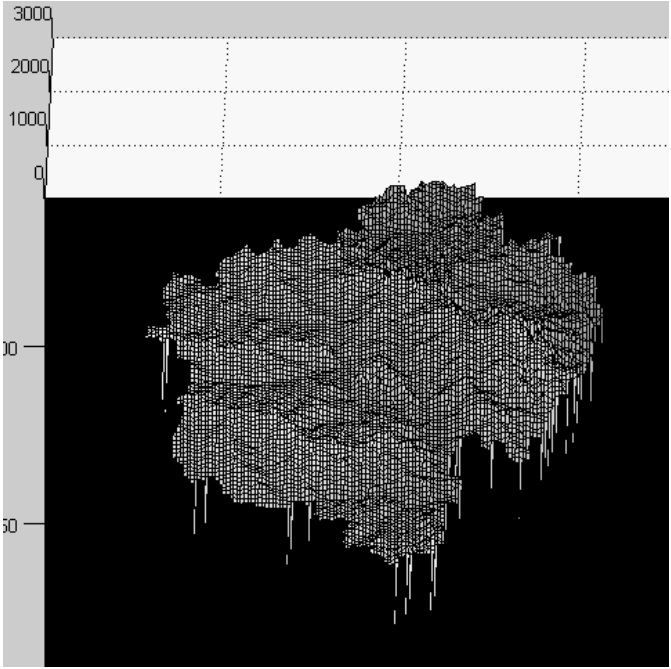

Figure 4.  A 70o camera view point of surface model with gray color scheme.

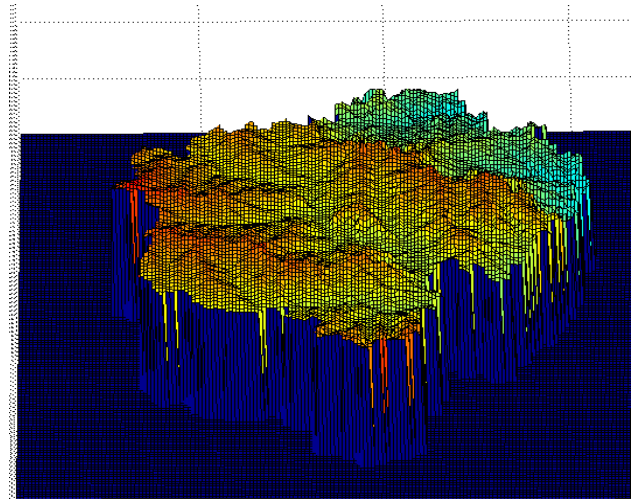

Figure 6.  A true color composite scheme (Atlas shader) 3D model.

## V. CONCLUSIONS AND FUTURE WORK

Some key processes for automated 3D geological surface modeling such as data acquisition, concatenation, 3D surface modeling and axes data managing have been presented.

The visualization experiments are done using data for Latur district. In the future work, we attempt to overlay real time map layers on this 3D surface model.

## ACKNOWLEDGEMENT

The authors are indebted to the National Remote Sensing Centre (NRSC), ISRO, India, for providing LULC digital data of Latur district, and to United States Geological Survey (USGS) for providing access to elevation data for Latur district.

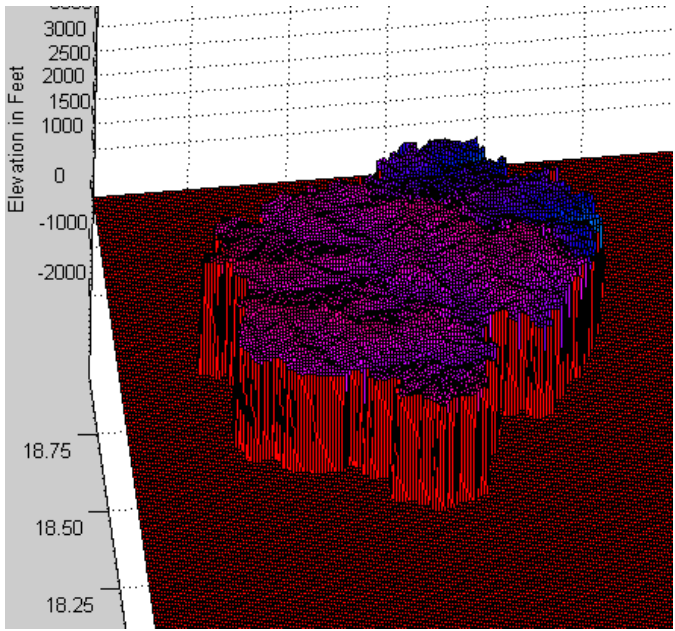

Figure 5.  A 45o camera view point of surface model with HSV color scheme.

AUTHORS PROFILE

**Kodge Bheemashankar G.** is a research scholar in department of studies and research in Computer Science of Swami Vivekanand College, Udgir Dist. Latur (MH) INDIA. He obtained MCM (Master in Computer Management) in 2004, M. Phil. in Computer Science in 2007 and registered for Ph.D. in Computer Science in 2008. His research areas of interests are GIS and Remote Sensing, Digital Image processing, Data mining and data warehousing. He is published 23 research papers in national, international Journals and proceedings conferences. Tel. +919923229672.

**Dr. P.S. Hiremath** is a Professor and Chairman, Department of P. G. Studies and Research in Computer Science, Gulbarga University, Gulbarga-585106 INDIA, He has obtained M.Sc. degree in 1973 and Ph.D. degree in 1978 in Applied Mathematics from Karnataka University, Dharwad. He had been in the Faculty of Mathematics and Computer Science of Various Institutions in India, namely, National Institute of Technology, Surathkal (1977-79), Coimbatore Institute of Technology, Coimbatore(1979-80), National Institute of Technology, Tiruchirapalli (1980-86), Karnatak University, Dharwad (1986-1993) and has been presently working as Professor of Computer Science in Gulbarga University, Gulbarga (1993 onwards). His research areas of interest are Computational Fluid Dynamics, Optimization Techniques, Image Processing and Pattern Recognition. He has published 142 research papers in peer reviewed International Journals and proceedings of conferences. Tel (off): +91 8472 263293, Fax: +91 8472 245927.